\documentclass[a4paper]{revtex4}
\usepackage{graphicx}
\usepackage{fancyhdr}
\usepackage{amsmath}
\usepackage{color}
\usepackage{amsmath,amssymb,mathrsfs}
\begin{document}


\title{Toward Higgs inflation in the MSSM}

%

\author{Takahiro Terada}
\affiliation{Department of Physics, The University of Tokyo, Tokyo 113-0033, Japan, \\
Deutsches Elektronen-Synchrotron (DESY), 22607 Hamburg, Germany}

\begin{abstract}
Adopting a recently proposed single-superfield framework of supergravity inflation, 
we consider large field inflationary models in which MSSM Higgs-like fields play the role of the inflaton.
In the simplest cases, the inflaton potential has a fractional power, which is different from that of the original Higgs inflation, and it can be tested by cosmological observations in near future.
We find difficulties in identifying the inflaton with the MSSM Higgses and discuss possible candidates of the inflaton.

\end{abstract}

\maketitle

\thispagestyle{fancy}


\section{Introduction}
 Higgs inflation~\cite{Bezrukov:2007ep} is a unique inflationary model that is based on the discovered scalar boson.  Its prediction of the spectral index and the tensor-to-scalar ratio is consistent with the Planck 2015 results~\cite{Ade:2015lrj}.  It is interesting to ask whether this scenario can be embedded in a theoretically motivated framework beyond the Standard Model (SM) of Particle Physics.  In this contribution, we study the possibility of large-field Higgs inflation in the Minimal Supersymmetric Standard Model (MSSM) taking into account the supergravity effects.
 
Inflation driven by the Higgs fields in the MSSM has been studied by several authors.
A small-field inflection-point inflation model was studied in Ref.~\cite{Chatterjee:2011qr} where a higher order Higgs term was added to the MSSM superpotential.
A large field model was studied in Ref.~\cite{Ibanez:2014kia, Ibanez:2014swa} where $\mu$ term as well as soft SUSY breaking mass terms are of order the inflation scale as high as $\mathcal{O}(10^{13})$ GeV.  The monodromy structure and string corrections are utilized in the model.
Here, we do not add higher order terms to the superpotential, and consider the possibility of Higgs inflation in the $\mathcal{N}=1$ supergravity framework.

The possibility of large field Higgs inflation in supergravity was studied and concluded that it does not occur due to the negative potential in the large field region~\cite{Einhorn:2009bh, BenDayan:2010yz}. Instead, Higgs inflation is realized in the Next-to-Minimal Supersymmetric Standard Model (NMSSM)~\cite{Einhorn:2009bh, BenDayan:2010yz, Lee:2010hj, Ferrara:2010yw, Ferrara:2010in} in which a singlet is added to the MSSM particle contents.  One of the reasons of the success is that the singlet can be naturally identified with the ``stabilizer field'' which is used for large field inflation in supergravity~\cite{Kawasaki:2000yn, Kallosh:2010ug}.

 Recently, Ketov and the author proposed a new method for making large field inflation possible in supergravity~\cite{Ketov:2014qha, Ketov:2014hya}.  In this approach, the stabilizer field is no longer required, and the positivity of the potential is restored by the shift-symmetric quartic term in the inflaton K\"{a}hler potential~\footnote{There are other similar but different approaches~\cite{Roest:2015qya, Linde:2015uga} that do not rely on the stabilizer field.
Despite the long history of studies on inflation in supergravity, these studies, including Refs.~\cite{Ketov:2014qha, Ketov:2014hya}, began just recently apart from a classic work~\cite{Goncharov:1983mw}.}. 
  It is, however, non-trivial whether the new method is applicable to the case of non-singlet fields (Higgs fields are charged electroweakly), and what kinds of inflaton potential are available.  We thus re-examine the possibility of Higgs inflation in the MSSM (\textit{i.e.} without the stabilizer singlet) by introducing higher dimensional terms in the Higgs K\"{a}hler potential.
 
\section{Large field inflation with Higgs(-like) fields} 
  
  Shift symmetry is a key to realize large field inflation in supergravity~\cite{Kawasaki:2000yn} since the $F$-term scalar potential has an exponential factor of the K\"{a}hler potential,
  \begin{align}
  V=e^{K}\left( K^{\bar{j}i}\left( W_i + K_i W\right) \left( \overline{W}_{\bar{j}}+K_{\bar{j}}\overline{W}\right) - 3|W|^2 \right),
  \end{align}
  where a bar ($\bar{\phantom{w}}$) denotes complex conjugation, and we use the reduced Planck unit, $M_{\text{P}}/\sqrt{8\pi}=1$.
  Since Higgs fields are charged, we consider shift transformation of a singlet combination of Higgs fields like $H_u H_d$ or its power $(H_u H_d)^n$~\cite{Nakayama:2010sk}, where  $H_u H_d$  is a short-hand notation for the SU(2) invariant contraction $H_u^{\text{t}}i\sigma_2 H_d=H_u^{+}H_d^{-}-H_u^0 H_d^0$.
  More generally, we consider the shift symmetry under the following transformation
 \begin{align}
 J(H_u H_d)\rightarrow J(H_u H_d)+ic, \label{shift}
 \end{align}
 where $J$ is an arbitrary holomorphic function, and $c$ is a constant.  We take $J(H_u H_d)=(\kappa H_u H_d )^n$ as a benchmark choice, where $\kappa$ is a constant.
 
  Consider the following K\"{a}hler potential and superpotential.
  \begin{align}
  K=&|H_u|^2+|H_d|^2+c\left( J ( H_u H_d ) +\bar{J}( \overline{H_u}\, \overline{H_d}) \right) \nonumber \\
  & +\frac{1}{2}\left( J ( H_u H_d ) +\bar{J}( \overline{H_u}\, \overline{H_d}) \right)^2 -\frac{\zeta}{4}\left( J ( H_u H_d ) +\bar{J}( \overline{H_u}\, \overline{H_d}) \right)^4, \label{Kpol} \\
  W=& \mu H_u H_d . \label{WMSSM}
  \end{align}
  The first two terms in the K\"{a}hler potential are responsible for the Higgs kinetic terms, and they break the shift symmetry~\eqref{shift}.
  The superpotential is that of the MSSM. 
  There are $2\times 4-3$(would-be Nambu-Goldstone bosons)$=5$ scalar degrees of freedom.  We truncate this theory to one with less degrees of freedom.  The mass of charged Higgs  $(\sim g/\sqrt{\kappa})$  is larger than the Hubble scale during inflation if we take not too large $\kappa$.  Then, charged Higgs is decoupled, and we neglect them hereafter.
  
  The K\"{a}hler potential is approximately constant along the quasi-$K$-flat direction,
  \begin{align}
  J(- H_u^0 H_d^0 ) + \bar{J} ( - \overline{H_u^0}\, \overline{H_d^0} )=0.
  \end{align}
  The $D$-term is constant along the $D$-flat direction,
  \begin{align}
  |H_u^0|=|H_d^0|.
  \end{align}
In the $K$- and $D$-flat direction, there is one scalar component, the inflaton.

In the field region satisfying $\left( |H_u^0|^2+|H_d^0|^2\right)|J'|^2\gg1$, the effect of the canonical terms on the kinetic term becomes negligible.  Taking $\kappa \gg 10^3$, we can also ignore its effects on the potential.  In the $D$-flat direction, the truncated theory becomes
  \begin{align}
  K\simeq c \left( \Phi+\bar{\Phi} \right) +\frac{1}{2}\left( \Phi+\bar{\Phi} \right)^2 - \frac{\zeta}{4}\left( \Phi+\bar{\Phi} \right)^4, 
  \end{align}
  where we defined a chiral superfield $\Phi=J(-H_u^0 H_d^0)$.
  Note that the above K\"{a}hler potential is same as that of single superfield framework with quartic stabilization~\cite{Ketov:2014qha}.
   The $K$-flat direction is now written as $\Phi+\bar{\Phi}=0$, \textit{i.e.} the imaginary direction, in which the inflaton rolls down. 
   The MSSM superpotential becomes 
  \begin{align}
  W=\mu J^{-1}(\Phi), \label{WJ}
  \end{align}
  where $J^{-1}$ is the inverse function of $J$.
  Thus, the potential in terms of the redefined inflaton superfield generically has a quite different shape from the original one.  This is a manifestation of the mechanism of running kinetic inflation~\cite{Nakayama:2010kt, Nakayama:2010sk}.
  
In our scenario, the inflationary scale ($\sim 10^{13}$ GeV in the case of typical chaotic inflation) is given by $\mu / \kappa$.
If we want to take $\mu$ as light as $\mathcal{O}(1)$ TeV, $\kappa$ has to be as small as $\mathcal{O}(10^{-10})$, but we need large $\kappa$ to neglect the canonical terms.
There is a possibility that the $\mu$ parameter is actually a dynamical field and its value changes after the inflation.  We do not pursue this scenario here since we do not include a singlet.
Next, let us consider the case $\mu\sim 10^{13}$ GeV.  First of all, SUSY should be broken also at this scale so that the $\mu$ parameter and soft SUSY breaking parameters cancel each other to reproduce the electroweak scale. That is, fine tuning is required.
In view of the heavy (125 GeV) Higgs mass, absence of SUSY signals at the LHC, and string landscape, this large $\mu$ term and fine tuning may not be a problem.
However, it is hard to control both effects of soft SUSY breaking terms and radiative corrections on the inflationary potential~\footnote{
Alternatively, one may consider the case where soft SUSY breaking terms drive inflation~\cite{Ibanez:2014kia, Buchmuller:2015oma}.
In our case, it leads to the potential $|V|\sim |(\mu^2 / \kappa ) J^{-1}(\Phi) |$.
This leads to the potential of fractional power $1/n$ when we choose $J=(\kappa H_u H_d)^n$.
}.
Also, a complete analysis would involve dynamics of both the Higgs-inflaton and SUSY breaking sectors.
We leave these issues for future investigation.

\begin{figure}[ht]
\centering
\includegraphics[width=80mm]{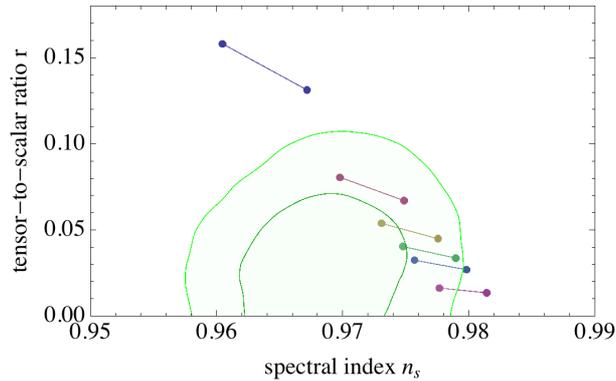}
\caption{Inflationary predictions of the fractional ($2/n$-th) power potentials.    The lines represent $n=1, 2, 3, 4, 5$, and 10, from top to bottom.  The left (right) points correspond to the $e$-folding number $N=50$ (60).
The light green contours are the constraint of Planck TT+lowP+BKP+lensing+BAO+JLA+$H_0$ (Fig. 21 in Ref.~\cite{Planck:2015xua}).  In this Figure, the correction to the $2/n$-th power term in Eq.~\eqref{V2/n} is not included.} \label{fig:ns_r}
\end{figure}

In the following, we switch to the possibilities of inflaton being MSSM Higgs-{\em like} fields.
We mean by ``MSSM Higgs-like'' that they have same quantum numbers as MSSM Higgses, and they have K\"{a}hler and superpotentials~\eqref{Kpol} and \eqref{WMSSM}, but they do not have large Yukawa couplings and their $\mu$ term may be larger than the electroweak scale.
Then, all of the above problems can be circumvented.
Let us specify the non-minimal K\"{a}hler function $J$ and see the resulting potential.
For simplicity, we take the monomial function $J$ of power $n$, $J(H_u H_d)=(\kappa H_u H_d)^n$, and we obtain the inflaton potential of asymptotic power $2/n$. 
\begin{align}
V=& \left| \frac{\mu}{\kappa}\right|^2 \left( (c^2-3)   \left( \frac{\chi}{\sqrt{2}}\right)^{\frac{2}{n}} +   \frac{c^2}{n^2} \left( \frac{\chi}{\sqrt{2}}\right)^{\frac{2}{n}-2}    \right), \label{V2/n}
\end{align}
where $\chi=\sqrt{2}\text{Im}\Phi$ is the inflaton.
Apparently, the potential diverges at the origin for $n> 1$, but this is an artifact of inappropriately extrapolating description in terms of the composite field $\Phi=J(-H_u^0 H_d^0 )$. 
In FIG.~\ref{fig:ns_r}, predictions of $2/n$-th power potential on the $(n_{\text{s}}, r)$-plane are shown with the Planck contours.  The lines correspond to purely quadratic, linear, $2/3$-th, $1/2$-th, $2/5$-th, and $1/5$-th power potentials ($2/n$-th power potentials with $n=1, 2, 3, 4, 5$ and 10), respectively.
The quadratic potential is disfavored by the latest observations, but the linear and fractional power potentials are within the $2\sigma$ contour of Planck constraints~\cite{Ade:2015lrj}.
In the Figure, we do not include the effects of subleading term in Eq.~\eqref{V2/n} because we have not take into account subleading terms in the function $J$, and also because it introduces a free parameter $c$.

We can also consider logarithmic K\"{a}hler potentials,
\begin{align}
K=& -a \ln \left( 1+\frac{1}{\sqrt{a}}\left( J(H_uH_d)+\bar{J}(\overline{H_u}\overline{H_d} )\right)  -\frac{1}{a} \left( \left| H_u \right| ^2 + \left| H_d \right|^2 \right)+\frac{\zeta}{a^2} \left( J(H_uH_d)+\bar{J}(\overline{H_u}\overline{H_d}) \right)^4   \right), \label{KHiggsLog}
\end{align}
where $a\geq 3$ is a constant.  This is invariant under the shift symmetry~\eqref{shift} up to the canonical terms $|H_u|^2$ and $|H_d|^2$.
We truncate the theory as above, and consider the $K$- and $D$-flat directions.
In the appropriate field range,
\begin{align}
\frac{ 1}{|J'|^2  } \left( 1+\frac{2}{\sqrt{a}} \left( H_u^0 H_d^0 J'+\overline{H_u^0}\overline{H_d^0}\bar{J}' \right)\right)   \ll |H_u^0|^2+|H_d^0|^2 \ll a, \label{kappaNeglectC}
\end{align}
we neglect the canonical terms. In terms of the inflaton superfield $\Phi=J(H_u H_d)$, the K\"{a}hler potential is the similar form to that in Ref.~\cite{Ketov:2014hya},
\begin{align}
K\simeq -a \ln \left( 1+\frac{1}{\sqrt{a}}\left(\Phi+\bar{\Phi}\right) +\frac{\zeta}{a^2}\left(\Phi+\bar{\Phi}\right)^4  \right).
\end{align}

Let us take the simple $J$ function again, $J=(\kappa H_u H_d)^n$.
The potential becomes
\begin{align}
V=&\left| \frac{\mu}{\kappa} \right|^2  \left( (a-3) \left( \frac{\chi}{\sqrt{2}}\right)^{\frac{2}{n}} +\frac{1}{n^2} \left( \frac{\chi}{\sqrt{2}}\right)^{\frac{2}{n}-2}   \right). \label{VHiggsALog}
\end{align}
The cases of $a\leq 2$ lead to negative potentials.
For $a=3$, the first term vanishes, and the potential has only the term with a non-positive power.
It is either a constant (for $n=1$), or a run-away type potential (for $n\geq 2$).
In the case of $a\geq 4$, the qualitative feature of the potential is same (asymptotically $2/n$-th power) as the potential~\eqref{V2/n} of the polynomial K\"{a}hler potential~\eqref{Kpol}.

\section{Conclusions}
We have found that large field inflation driven by MSSM Higgs inflaton is non-trivial to achieve in supergravity.
Even if the specifically chosen K\"{a}hler potential and fine-tuning to reproduce the electroweak scale are accepted, the soft masses and radiative corrections affect the inflaton potential.
We then loosened the requirements and consider some Higgs-like fields.
The resultant potential is different from the plateau-type potential of the original Higgs inflation, and it is derived from the inverse function of an arbitrary holomorphic function in the effective superpotential (see Eq.~\eqref{WJ}).
With simplifying assumptions stated above, we have fractional $2/n$-th power potentials.
The simplest case ($n=1$; quadratic potential) is now disfavored by the Planck data, but it is interesting that next-to-simplest cases ($n\geq 2$) can be tested in near future.

Let us finally discuss what the ``MSSM Higgs-like fields'' are.
A primary candidate is a GUT Higgs, which requires some modifications of the model considered here.
It would explain the large inflationary scale with less tuning.
One could also consider a Kaluza-Klein excitation of the MSSM Higgs fields, which can be as heavy as the inflationary scale.
Moreover, in a superstring setup with multiple Higgs doublets, a small $\mu$-term can be obtained without fine tuning~\cite{Abe:2015uma}.  In this model, the $\mu$-terms for other Higgses can be in the range between $\mathcal{O}(10^{12})$ GeV and $\mathcal{O}(10^{18})$ GeV, so they could be used as the inflaton if the K\"{a}hler potential can be modified as Eqs.~\eqref{Kpol} or \eqref{KHiggsLog}.
In any case, one has to ensure small enough couplings not to break the shift symmetry significantly.
A large $\kappa$ suppresses couplings in terms of the canonically normalized field $\Phi$ and improves the description.  Taking $\kappa\sim10^3$ or $10^5$, one can have a GUT or Planck scale $\mu$ term, which are compatible with the above candidates.
Studies on these possibilities are to be done elsewhere.

\begin{acknowledgments}
I am grateful to T.~Kitahara, K.~Mukaida, S.~Shirai, and M.~Takimoto for useful discussion.
I also thank Y.~Tatsuta for introducing me Ref.~\cite{Abe:2015uma} and discussion on it.
I am supported by a Grant-in-Aid for JSPS Fellows, and a JSPS Grant-in-Aid for Scientific Research No.~2610619.
\end{acknowledgments}

\bigskip 

\end{document}